\documentclass[twocolumn,showpacs,preprintnumbers,amsmath,amssymb,superscriptaddress]{revtex4}

\usepackage{graphicx}
\usepackage{dcolumn}
\usepackage{bm}

\begin{document}

\title{Measurement of Day and Night Neutrino Energy Spectra at SNO and\\
Constraints on Neutrino Mixing Parameters}

%
\newcommand{\ubc}{Department of Physics and Astronomy, University of 
British Columbia, Vancouver, BC V6T 1Z1 Canada}
\newcommand{\bnl}{Chemistry Department, Brookhaven National 
Laboratory,  Upton, NY 11973-5000}
\newcommand{\carleton}{Carleton University, Ottawa, Ontario K1S 5B6 Canada}
\newcommand{\uog}{Physics Department, University of Guelph,  
Guelph, Ontario N1G 2W1 Canada}
\newcommand{\lu}{Department of Physics and Astronomy, Laurentian 
University, Sudbury, Ontario P3E 2C6 Canada}
\newcommand{\lbnl}{Institute for Nuclear and Particle Astrophysics and 
Nuclear Science Division, Lawrence Berkeley National Laboratory, Berkeley, CA 94720}
\newcommand{\lanl}{Los Alamos National Laboratory, Los Alamos, NM 87545}
\newcommand{\oxford}{Department of Physics, University of Oxford, 
Denys Wilkinson Building, Keble Road, Oxford, OX1 3RH, UK}
\newcommand{\penn}{Department of Physics and Astronomy, University of 
Pennsylvania, Philadelphia, PA 19104-6396}
\newcommand{\queens}{Department of Physics, Queen's University, 
Kingston, Ontario K7L 3N6 Canada}
\newcommand{\uw}{Center for Experimental Nuclear Physics and Astrophysics, 
and Department of Physics, University of Washington, Seattle, WA 98195}
\newcommand{\triumf}{TRIUMF, 4004 Wesbrook Mall, Vancouver, BC V6T 2A3, Canada}
\newcommand{\ralsuss}{Rutherford Appleton Laboratory, Chilton, Didcot, 
Oxon, OX11 0QX, and University of Sussex, Physics and Astronomy Department, 
Brighton BN1 9QH, UK}
\newcommand{\uci}{Department of Physics, University of California, 
Irvine, CA 92717}
\newcommand{\aecl}{Atomic Energy of Canada, Limited, Chalk River Laboratories, 
Chalk River, ON K0J 1J0, Canada}
\newcommand{\nrc}{National Research Council of Canada, Ottawa, ON K1A 0R6, Canada}
\newcommand{\princeton}{Department of Physics, Princeton University, 
Princeton, NJ 08544}
\newcommand{\birkbeck}{Birkbeck College, University of London, Malet Road, 
London WC1E 7HX, UK}

\affiliation{	\aecl	}
\affiliation{	\ubc	}
\affiliation{	\bnl	}
\affiliation{	\uci	}
\affiliation{	\carleton	}
\affiliation{	\uog	}
\affiliation{	\lu	}
\affiliation{	\lbnl	}
\affiliation{	\lanl	}
\affiliation{	\nrc	}
\affiliation{	\oxford	}
\affiliation{	\penn	}
\affiliation{	\princeton	}
\affiliation{	\queens	}
\affiliation{	\ralsuss	}
\affiliation{	\triumf	}
\affiliation{	\uw	}

\author{	Q.R.~Ahmad	}\affiliation{	\uw	}
\author{	R.C.~Allen	}\affiliation{	\uci	}
\author{	T.C.~Andersen	}\affiliation{	\uog	}			
\author{	J.D.~Anglin	}\affiliation{	\nrc	}			
\author{	J.C.~Barton	}\altaffiliation[Permanent Address: ]{\birkbeck}	\affiliation{	\oxford	}		
\author{	E.W.~Beier	}\affiliation{	\penn	}			
\author{	M.~Bercovitch	}\affiliation{	\nrc	}			
\author{	J.~Bigu	}\affiliation{	\lu	}			
\author{	S.D.~Biller	}\affiliation{	\oxford	}			
\author{	R.A.~Black	}\affiliation{	\oxford	}			
% ...

\collaboration{SNO Collaboration}
\noaffiliation

\date{\today}

\begin{abstract}
The Sudbury Neutrino Observatory (SNO) has measured day and night
solar neutrino energy spectra and rates. For charged current events,
assuming an undistorted $^8$B spectrum, the night minus day rate is
$14.0\% \pm 6.3\% ^{+1.5}_{-1.4}\%$ of the average rate.  If the total
flux of active neutrinos is additionally constrained to have no
asymmetry, the $\nu_e$ asymmetry is found to be $7.0\% \pm 4.9\%
^{+1.3}_{-1.2}\%$. A global solar neutrino analysis in terms of
matter-enhanced oscillations of two active flavors strongly favors the
Large Mixing Angle (LMA) solution.
\end{abstract}

\pacs{26.65.+t, 14.60.Pq, 13.15.+g, 95.85.Ry}
\maketitle

The Sudbury Neutrino Observatory (SNO) has provided strong evidence
that neutrinos change flavor as they propagate from the core of the
Sun~\cite{bib:SNONC, bib:1stPRL}, independently of solar model flux
predictions.  This flavor conversion can be explained by neutrino
oscillation models based on flavor mixing.  For some values of the
mixing parameters, spectral distortions and a measurable dependence on
solar zenith angle are expected~\cite{bib:theo1,bib:theo2,bib:theo3}.
The latter might be caused by interaction with the matter of the Earth
(the MSW effect) and would depend not only on oscillation parameters
and neutrino energy, but also on the path length and $e^-$ density
through the Earth.  This Letter presents SNO's first measurements of
day and night neutrino energy spectra, and constraints on allowed
neutrino mixing parameters.

The data reported here were recorded between November 2, 1999 and May
28, 2001 UTC.  The total livetimes for day and night are 128.5 and
177.9 days, respectively.  The time-averaged inverse-square distance
to the Sun $\langle(\frac{1AU}{R})^2\rangle$ was 1.0002 (day) and
1.0117 (night).  During the development of this analysis, the data
were partitioned into two sets of approximately equal livetime (split
at July 1, 2000), each having substantial day and night components.
Analysis procedures were refined during the analysis of Set 1 and
fixed before Set 2 was analysed. The latter thus served as an unbiased
test.  Unless otherwise stated, the analysis presented in this paper
is for the combined data set.

The data reduction in \cite{bib:SNONC} was used here.  For each event,
the number, pattern, and timing of the hit photomultiplier tubes
(PMTs) were used to reconstruct effective recoil electron kinetic
energy $T_{\emph{eff}}$, radial position $R$, and scattering angle
$\theta_{\odot}$ with respect to the Sun-Earth direction.  The charged
current (CC), elastic scattering (ES) and neutral current (NC)
reactions each have characteristic probability density functions
(pdfs) of $T_{\emph{eff}}, R$, and $\theta_{\odot}$.  A maximum
likelihood fit of the pdfs to the data determined the flux from each
of these reactions.

The measured night and day fluxes $\phi_{N}$ and $\phi_{D}$ were used
to form the asymmetry ratio for each reaction: $\mathcal{A} = 2
(\phi_{N} - \phi_{D})/(\phi_{N} + \phi_{D})$.  The CC interaction is
sensitive only to $\nu_e$.  The NC interaction is equally sensitive to
all active neutrino flavors, so active-only neutrino models predict
$\mathcal{A}_{NC} = 0$ ~\cite{sterile}.  The same models allow
$\mathcal{A}_{CC} \not=0$.  The ES reaction has additional
contributions from $\nu_{\mu\tau}$ leading to a reduction in its
sensitivity to $\nu_e$ asymmetries.

SNO used calibration sources~\cite{NIM} to constrain variations in
detector response~\cite{bib:companion} that can lead to day-night
asymmetries.  A $^{16}$N source~\cite{N16}, which produces $6.1$-MeV
gamma rays, revealed a 1.3\% per year drift in the energy scale.  Due
to seasonal variation in day and night livetime, this drift can create
an artificial asymmetry.  The analysis corrected for this drift and a
systematic uncertainty was assigned using worst-case drift
models. Gamma rays from the $^{16}$N source were also used to
constrain directional dependences in SNO's response.

\begin{figure}
\begin{center}
\includegraphics[width=3.6in]{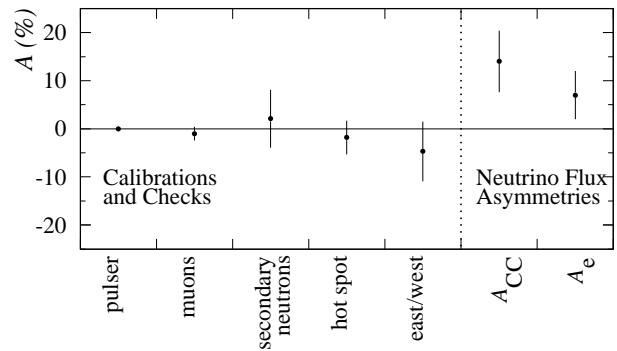}
\caption{\label{fig:DNcal} Various event classes used to determine
systematic differences between day and night measurements.  Also shown
are measured asymmetries on the CC flux, and on the electron neutrino
flux derived from the CC, ES, and NC rates when the total neutrino
flux is constrained to have zero asymmetry.  }
\end{center}
\end{figure}

A set of signals that are continuously present in the detector was
used to probe possible diurnal variations in detector response.  The
detector was triggered at 5 Hz with a pulser, verifying livetime
accounting. Muons provide an almost constant signal and, through
interactions with D$_2$O, produce secondary neutrons.  After applying
a cut to remove bursts with high neutron multiplicity, these
muon-induced neutrons were used to limit temporal variations in
detector response.  A more sensitive study focused on a solitary point
of high background radioactivity, or ``hot spot'', on the upper
hemisphere of the SNO acrylic vessel, apparently introduced during
construction.  Its event rate was stable and sufficient to make an
excellent test of diurnal variations.  It also provides a sensitive
test for changes in reconstruction.  A limit of 3.5\% on the hot spot
rate asymmetry was determined, which because of its steeply falling
energy spectrum constrained the day and night energy scales to be the
same within 0.3\%.  An east/west division of the neutrino data based
on the Sun's position should show no rate variations from matter
effects.  As expected, the CC rates for east and west data were
consistent.  The rate asymmetries for each test are shown in
Fig.~\ref{fig:DNcal}.

\begin{figure}
\begin{center}
\includegraphics[width=3.6in]{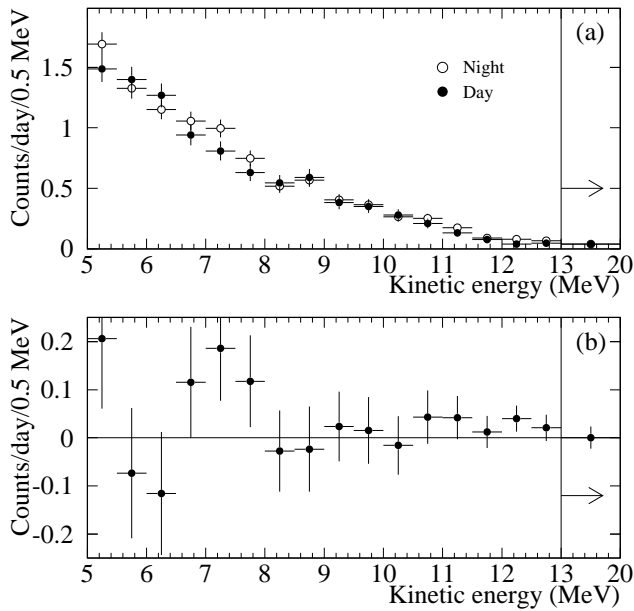}
\caption{\label{fig:DNspectra} (a) Energy spectra for day and night.
All signals and backgrounds contribute. The final bin extends from
13.0 to 20.0 MeV.  (b) Difference, \emph{night - day}, between the
spectra.  The day rate was $9.23 \pm 0.27$ events/day, and the night
rate was $9.79 \pm 0.24$ events/day.}
\end{center}
\end{figure}

\begin{figure}
\begin{center}
\includegraphics[width=3.6in]{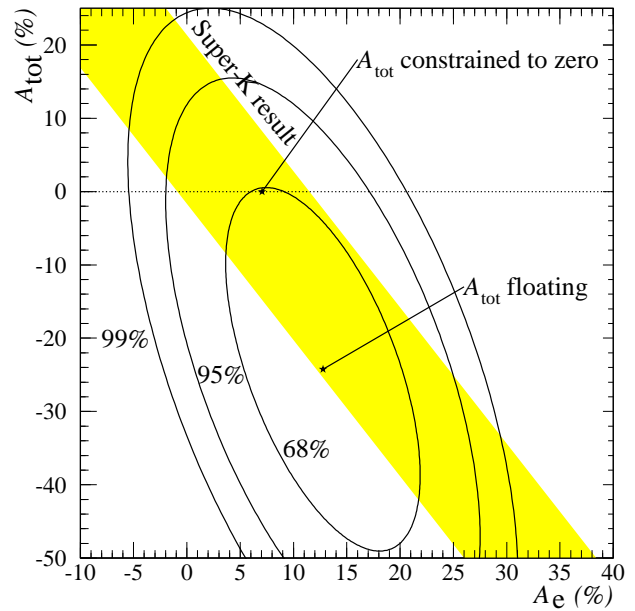}
\caption{\label{fig:DNcorr} Joint probability contours for
$\mathcal{A}_{tot}$ and $\mathcal{A}_{e}$.  The points indicate the
results when $\mathcal{A}_{tot}$ is allowed to float and when it is
constrained to zero.  The diagonal band indicates the 68\% joint
contour for the Super-K $\mathcal{A}_{ES}$ measurement.  }
\end{center}
\end{figure}

\begin{table*}
\caption{\label{tab:sigex} The results of signal extraction, assuming
an undistorted $^8$B spectrum.  The systematic uncertainties (combined
set) include a component that cancels in the formation of the
$\mathcal{A}$.  Except for the dimensionless $\mathcal{A}$, the units
are $10^6$~cm$^{-2}$~s$^{-1}$.  Flux values have been rounded, but the
asymmetries were calculated with full precision.
}
\begin{ruledtabular}
\begin{tabular}{c|cc|cc|cc|c}
 & \multicolumn{2}{c}{Set 1} & \multicolumn{2}{c}{Set 2} &
\multicolumn{2}{c}{Combined} &   $\mathcal{A} (\%)$ \\ 
signal & $\phi_{D}$ & $\phi_{N}$ & $\phi_{D}$ & $\phi_{N}$ &
$\phi_{D}$ & $\phi_{N}$ & \\  
\hline
CC & $1.53\pm0.12$ & $1.95\pm0.10$ & $1.69\pm0.12$ &$1.77\pm0.11$ &
$1.62\pm0.08\pm 0.08$  & $1.87\pm 0.07\pm 0.10$ & $+14.0 \pm~6.3
^{+1.5}_{-1.4}$ \\ 
ES & $2.91\pm0.52$ & $1.59\pm0.38$ & $2.35\pm0.51$ &$2.88\pm0.47$ &
$2.64\pm0.37\pm 0.12$ & $2.22 \pm0.30\pm 0.12$ &  $-17.4 \pm 19.5
^{+2.4}_{-2.2}$ \\ 
NC & $7.09\pm0.97$ & $3.95\pm0.75$ & $4.56\pm0.89$ &$5.33\pm0.84$ &
$5.69\pm0.66\pm 0.44$  & $4.63\pm0.57\pm 0.44$  &  $-20.4 \pm 16.9
^{+2.4}_{-2.5}$ \\ 
\end{tabular}
\end{ruledtabular}
\end{table*}

\begin{table}
\caption{\label{tab:syserrs} Effect of systematic uncertainties on
$\mathcal{A}~(\%)$.  For presentation, uncertainties have been
symmetrized and rounded.}
\begin{ruledtabular}
\begin{tabular}{c|c|c|c}
Systematic & $\delta \mathcal{A}_{CC} $ & $\delta \mathcal{A}_{ES} $ &
$\delta \mathcal{A}_{NC} $ \\ 
\hline
Long-term energy scale drift       & 0.4 & 0.5 & 0.2 \\
Diurnal energy scale variation     & 1.2 & 0.7 & 1.6 \\
Directional energy scale var.      & 0.2 & 1.4 & 0.3 \\
Diurnal energy resolution var.     & 0.1 & 0.1 & 0.3 \\
Directional energy resolution var. & 0.0 & 0.1 & 0.0 \\
Diurnal vertex shift var.              & 0.5 & 0.6 & 0.7 \\
Directional vertex shift var. & 0.0 & 1.1 & 0.1 \\
Diurnal vertex resolution var.     & 0.2 & 0.7 & 0.5 \\
Directional angular recon. var.    & 0.0 & 0.1 & 0.1 \\
PMT $\beta$-$\gamma$ background    & 0.0 & 0.2 & 0.5 \\
AV+H$_2$O $\beta$-$\gamma$ bkgd.          & 0.0 & 0.6 & 0.2 \\
D$_2$O $\beta$-$\gamma$, neutrons bkgd.   & 0.1 & 0.4 & 1.2 \\
External neutrons bkgd.                 & 0.0 & 0.2 & 0.4 \\
Cut acceptance                          & 0.5 & 0.5 & 0.5 \\
\hline
Total                              & 1.5 & 2.4 & 2.4 \\
\end{tabular}
\end{ruledtabular}

\caption{\label{tab:electron} Measurement of the $\phi_e$ and
$\phi_{tot}$ asymmetry for various constraints.  All analyses assume
an undistorted $^8$B spectrum.  }
\begin{ruledtabular}
\begin{tabular}{l|c}
Constraints & Asymmetry (\%)\\  
\hline
a) no additional constraint & $\mathcal{A}_{CC} = 14.0 \pm 6.3
^{+1.5}_{-1.4}$\\ 
			    & $\mathcal{A}_{NC} = -20.4 \pm 16.9
^{+2.4}_{-2.5}$\\ 
 & (see text for correlations)\\
\hline
b) $\phi_{ES} = (1-\epsilon) \phi_{e} + \epsilon \phi_{tot}$ &
$\mathcal{A}_{e} = 12.8 \pm 6.2 ^{+1.5}_{-1.4}$\\ 
&  $\mathcal{A}_{tot} = -24.2 \pm 16.1 ^{+2.4}_{-2.5}$\\ 
& correlation = -0.602\\
\hline
c) $\phi_{ES} = (1-\epsilon) \phi_{e} + \epsilon \phi_{tot}$ & \\
$~~~\mathcal{A}_{tot} = 0$& $\mathcal{A}_{e} = 7.0 \pm 4.9 ^{+1.3}_{-1.2}$\\
\hline
d) $\phi_{ES} = (1-\epsilon) \phi_{e} + \epsilon \phi_{tot}$ &
$\mathcal{A}_e(SK) = 5.3 \pm 3.7 ^{+2.0}_{-1.7}$\\ 
$~~~\mathcal{A}_{tot} = 0$& (derived from SK $\mathcal{A}_{ES}$ \\
$~~~\mathcal{A}_{ES}(SK) = 3.3\% \pm 2.2\% ^{+1.3}_{-1.2}\%$ & and SNO
total $^8$B flux)\\ 
\end{tabular}
\end{ruledtabular}
\end{table}

Backgrounds were subtracted separately for day and night as part of
the signal extraction. The results were normalized for an Earth-Sun
distance of 1 AU, yielding the results in Table \ref{tab:sigex}.  Day
and night fluxes are given separately for data Sets 1 and 2, and for
the combined data.  A $\chi^2$ consistency test of the six measured
fluxes between Sets 1 and 2 yielded a chance probability of 8\%.  A
similar test done directly on the three asymmetry parameters gave a
chance probability of 2\%.  No systematic has been identified, in
either signal or background regions, that would suggest that the
differences between Set 1 and Set 2 are other than a statistical
fluctuation.  For the combined analysis, $\mathcal{A}_{CC}$ is
$+2.2\sigma$ from zero, while $\mathcal{A}_{ES}$ and
$\mathcal{A}_{NC}$ are $-0.9\sigma$ and $-1.2\sigma$ from zero,
respectively.  Note that $\mathcal{A}_{CC}$ and $\mathcal{A}_{NC}$ are
strongly statistically anti-correlated ($\rho=-0.518$), while
$\mathcal{A}_{CC}$ and $\mathcal{A}_{ES}$ ($\rho=-0.161$) and
$\mathcal{A}_{ES}$ and $\mathcal{A}_{NC}$ ($\rho=-0.106$) are
moderatedly anti-correlated.  Table \ref{tab:syserrs} gives the
systematic uncertainties on the asymmetry parameters.  The day and
night energy spectra for all accepted events are shown in
Fig.~\ref{fig:DNspectra}.

Table \ref{tab:electron} (a) shows the results for $\mathcal{A}_e$
derived from the CC day and night rate measurements, i.e.,
$\mathcal{A}_e = \mathcal{A}_{CC}$.  The day and night flavor contents
were then extracted by changing variables to $\phi_{CC} = \phi_{e}$,
$\phi_{NC} = \phi_{tot} = \phi_{e}+\phi_{\mu\tau}$ and $\phi_{ES} =
\phi_{e} + \epsilon \phi_{\mu\tau}$, where $\epsilon \equiv 1/6.48$ is
the ratio of the average ES cross sections above 5~MeV for
$\nu_{\mu\tau}$ and $\nu_e$.  Table \ref{tab:electron} (b) shows the
asymmetries of $\phi_e$ and $\phi_{tot}$ with this additional
constraint from the ES rate measurements.  This analysis allowed for
an asymmetry in the total flux of $^8$B neutrinos (non-zero
$\mathcal{A}_{tot}$), with the measurements of $\mathcal{A}_{e}$ and
$\mathcal{A}_{tot}$ having a strong anti-correlation.
Fig.~\ref{fig:DNcorr} shows the $\mathcal{A}_e$
vs. $\mathcal{A}_{tot}$ joint probability contours.  Forcing
$\mathcal{A}_{tot} = 0$, as predicted by active-only models, yielded
the result in Table \ref{tab:electron} (c) of $\mathcal{A}_{e} = 7.0\%
\pm 4.9\%~\mathrm{(stat.)  ^{+1.3}_{-1.2}}\%~\mathrm{(sys.)}$.

The Super-Kamiokande (SK) collaboration measured $\mathcal{A}_{ES}(SK)
= 3.3\% \pm
2.2\%~\mathrm{(stat.)}^{+1.3}_{-1.2}\%~\mathrm{(sys.)}$~\cite{bib:SuperK}.
The ES measurement includes a neutral current component, which reduces
the asymmetry for this reaction relative to
$\mathcal{A}_{e}$~\cite{Robustsig}.  $\mathcal{A}_{ES}(SK)$ may be
converted to an equivalent electron flavor asymmetry using the total
neutrino flux measured by SNO, yielding $\mathcal{A}_{e}(SK)$ (Table
\ref{tab:electron} (d)).  This value is in good agreement with SNO's
direct measurement of $\mathcal{A}_e$, as seen in
Fig.~\ref{fig:DNcorr}.
 
\begin{figure*}
\begin{center}
\includegraphics[width=3.5in]{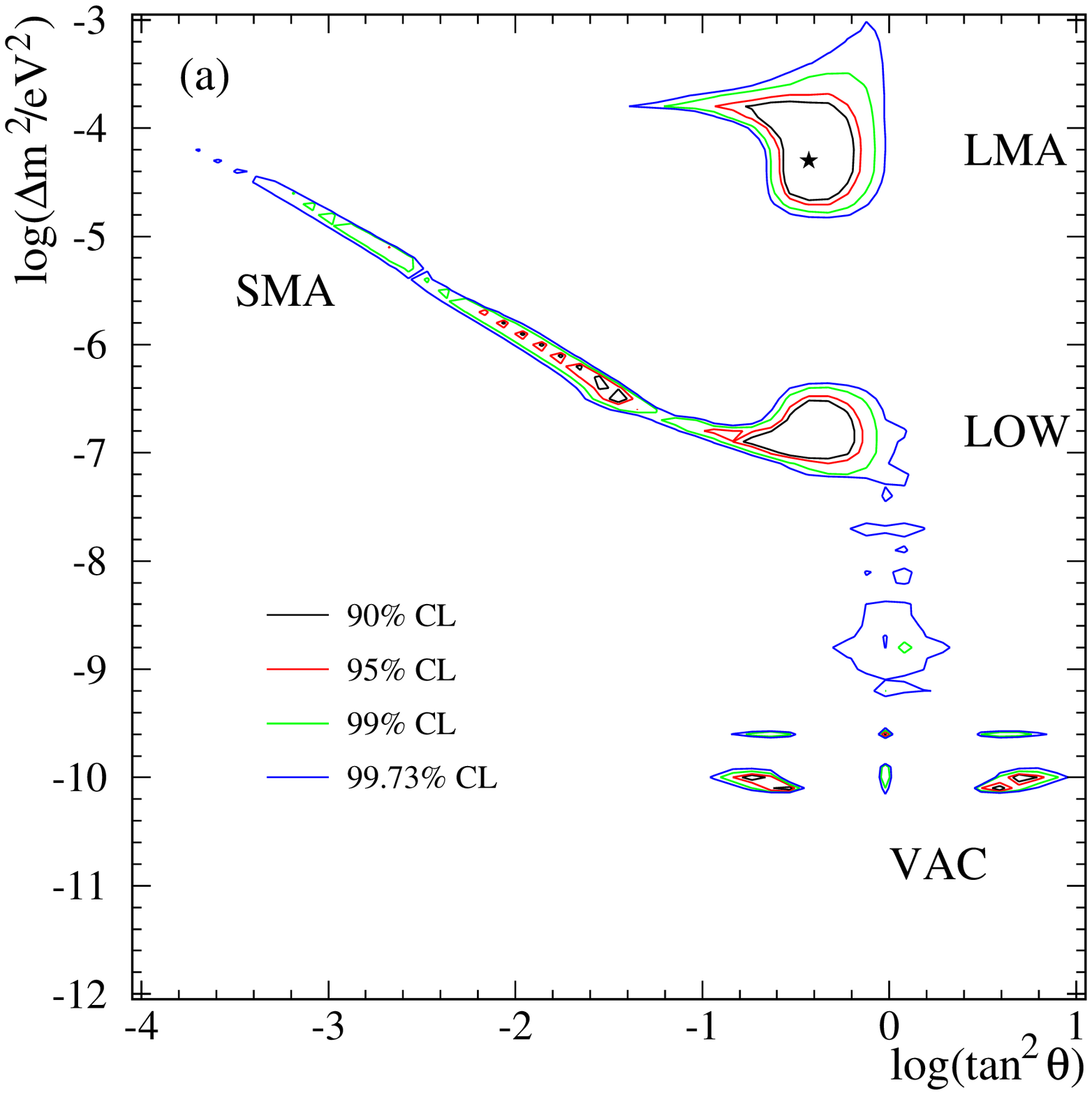}
\includegraphics[width=3.5in]{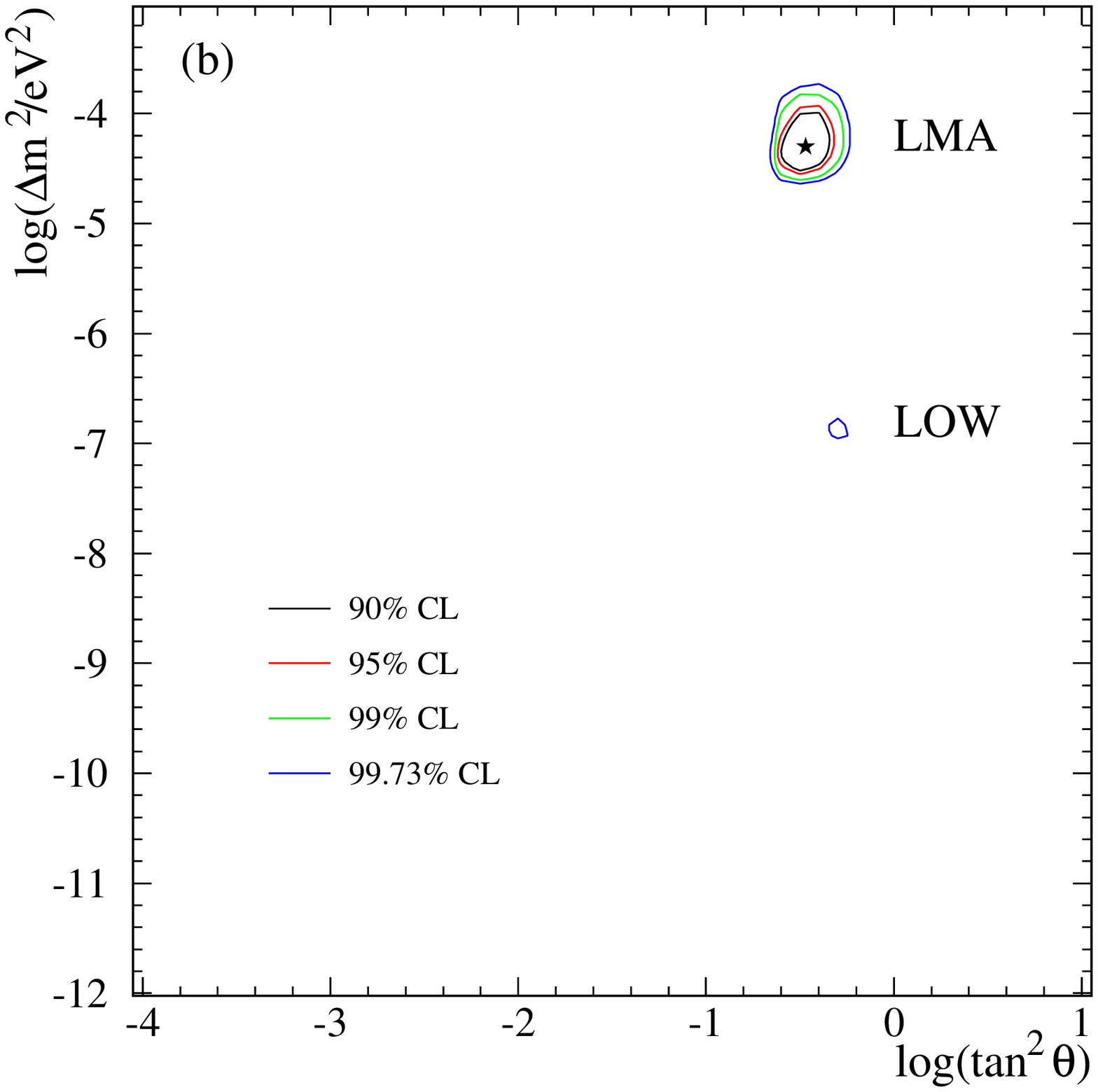}
\caption{\label{fig:MSWsno}
Allowed regions of the MSW plane determined  by a $\chi^2$ fit to (a) 
SNO day and night energy spectra and (b) with additional
experimental and solar model data.  The star indicates the best fit.
See text for details.} 
\end{center}
\end{figure*}

SNO's day and night energy spectra (Fig.~\ref{fig:DNspectra}) have
also been used to produce MSW exclusion plots and limits on neutrino
flavor mixing parameters.  MSW oscillation models~\cite{bib:MSW}
between two active flavors were fit to the data.  For simplicity, only
the energy spectra were used in the fit, and the radial $R$ and
direction $\cos\theta_{\odot}$ information was omitted.  This
procedure preserves most of the ability to discriminate between
oscillation solutions.  A model was constructed for the expected
number of counts in each energy bin by combining the neutrino spectrum
\cite{bib:8B}, the survival probability, and the cross sections
\cite{bib:xsec} with SNO's response functions~\cite{bib:companion}.
For this analysis, the dominant systematics are those for the combined
fluxes, as detailed in Refs. \cite{bib:SNONC} and
\cite{bib:companion}, and not the diurnal systematics of
Table~\ref{tab:syserrs}.

There are 3 free parameters in the fit: the total $^8$B flux
$\phi_{B}$, the difference $\Delta m^2$ between the squared masses of
the two neutrino mass eigenstates, and the mixing angle $\theta$. The
flux of higher energy neutrinos from the solar \emph{hep} reaction was
fixed at $9.3 \times 10^{3}$~cm$^{-2}$~s$^{-1}$~\cite{bib:SSM1}.
Contours were generated in $\Delta m^2$ and $\tan^2 \theta$ for
$\Delta \chi^2(c.l.) = 4.61~(90\%), 5.99~(95\%), 9.21~(99\%)$, and
$11.83~(99.73\%)$.  Fig.~\ref{fig:MSWsno}(a) shows allowed mixing
parameter regions using only SNO data with no additional experimental
constraints or inputs from solar models.  By including flux
information from the Cl~\cite{bib:Cl} and Ga
experiments~\cite{bib:Sage, bib:Sage3, bib:Gallex, bib:GALLEX2,
bib:Gallex3}, the day and night spectra from the SK
experiment~\cite{bib:SuperK}, along with solar model predictions for
the more robust $pp$, $pep$ and $^7$Be neutrino
fluxes~\cite{bib:SSM1}, the contours shown in Fig. \ref{fig:MSWsno}(b)
were produced.  This global analysis strongly favors the Large Mixing
Angle (LMA) region (see Table \ref{tab:chisq}), and $\tan^2 \theta$
values $<1$.  While the absolute chi-squared per degree of freedom is
not particularly large for the LOW solution, the difference between
chi-squared values still reflects the extent to which one region of
MSW parameter space is favored compared to another.  Repeating the
global analysis using the total SNO energy spectrum instead of
separate day and night spectra gives nearly identical results.

In summary, SNO has measured the day-night asymmetries of the CC, NC,
and ES reaction rates.  From these results the first direct
measurements of the day-night asymmetries in the $\nu_e$ flux and the
total $\nu$ flux from the Sun have been deduced.  A global fit to
SNO's day and night energy spectra and data from other solar neutrino
experiments strongly favors the LMA solution in a 2-flavor MSW
neutrino oscillation analysis.

This research was supported by: Canada: NSERC, Industry Canada, NRC,
Northern Ontario Heritage Fund Corporation, Inco, AECL, Ontario Power
Generation; US: Dept. of Energy; UK: PPARC.  We thank the SNO
technical staff for their strong contributions.

\begin{table}[h]
\caption{\label{tab:chisq} Best fit points in the MSW plane for global
MSW analysis using all solar neutrino data.  $\phi_{B}$ is the
best-fit $^8$B flux for each point, and has units of
$10^6$~cm$^{-2}$~s$^{-1}$. $\Delta m^2$ has units of eV$^2$.
$\mathcal{A}_e$ is the predicted asymmetry for each point.}
\begin{ruledtabular}
\begin{tabular}{c|cccccc}
Region & $\chi^2_{min}$/dof & $\phi_{B}$ & $\mathcal{A}_e (\%)$ &
$\Delta m^2$ & $\tan^2 \theta$ & c.l.(\%)\\
\hline
LMA &  57.0/72 & 5.86 & 6.4 & $5.0 \times 10^{-5}$ & 0.34 &  --- \\
LOW &  67.7/72 & 4.95 & 5.9 & $1.3 \times 10^{-7}$ & 0.55 &  99.5 \\
\end{tabular}
\end{ruledtabular}
\end{table}


\begin{thebibliography}{18}
\expandafter\ifx\csname natexlab\endcsname\relax\def\natexlab#1{#1}\fi
\expandafter\ifx\csname bibnamefont\endcsname\relax
  \def\bibnamefont#1{#1}\fi
\expandafter\ifx\csname bibfnamefont\endcsname\relax
  \def\bibfnamefont#1{#1}\fi
\expandafter\ifx\csname citenamefont\endcsname\relax
  \def\citenamefont#1{#1}\fi
\expandafter\ifx\csname url\endcsname\relax
  \def\url#1{\texttt{#1}}\fi
\expandafter\ifx\csname urlprefix\endcsname\relax\def\urlprefix{URL }\fi
\providecommand{\bibinfo}[2]{#2}
\providecommand{\eprint}[2][]{\url{#2}}

\bibitem[{\citenamefont{Ahmad {\em et~al.}}(2002)}]{bib:SNONC}
\bibinfo{auth     or}{\bibfnamefont{Q.~R.} \bibnamefont{Ahmad}}
  \bibnamefont{{\em et~al.}}, \bibinfo{journal}{Phys. Rev. Lett.}
  (\bibinfo{year}{2002}), \bibinfo{note}{SNO NC PRL, submitted for
publication}.

\bibitem[{\citenamefont{Ahmad {\em et~al.}}(2001)}]{bib:1stPRL}
\bibinfo{author}{\bibfnamefont{Q.~R.} \bibnamefont{Ahmad}}
  \bibnamefont{{\em et~al.}}, \bibinfo{journal}{Phys. Rev. Lett.}
  \textbf{\bibinfo{volume}{87}}, \bibinfo{pages}{071301}
  (\bibinfo{year}{2001}).

\bibitem[{\citenamefont{{S.~P. Mikheyev and A.~Y. Smirnov}}(1986)}]{bib:theo1}
\bibinfo{author}{\bibnamefont{{S.~P. Mikheyev and A.~Y. Smirnov}}}, in
  \emph{\bibinfo{booktitle}{'86 Massive Neutrinos in Astrophysics and in
  Particle Physics, Proceedings of the Moriond Workshop}}, edited by
  \bibinfo{editor}{\bibnamefont{{O. Fackler and J. Tran Thanh Van}}}
  (\bibinfo{year}{1986}), vol. \bibinfo{volume}{Editions Fronti\`eres,
  Gif-sur-Yvette, 1986}, p. \bibinfo{pages}{335}.

\bibitem[{\citenamefont{{A.~J. Baltz and J. Weneser}}(1988)}]{bib:theo2}
\bibinfo{author}{\bibnamefont{{A.~J. Baltz and J. Weneser}}},
  \bibinfo{journal}{Phys. Rev. D} \textbf{\bibinfo{volume}{37}},
  \bibinfo{pages}{3364} (\bibinfo{year}{1988}).

\bibitem[{\citenamefont{{M.~C. Gonzalez-Garcia, C. Pe\~na-Garay and A.~Y.
  Smirnov}}(2001)}]{bib:theo3}
\bibinfo{author}{\bibnamefont{{M.~C. Gonzalez-Garcia, C. Pe\~na-Garay and A.~Y.
  Smirnov}}}, \bibinfo{journal}{Phys. Rev. D} \textbf{\bibinfo{volume}{63}},
  \bibinfo{pages}{113004} (\bibinfo{year}{2001}).

\bibitem[{ste()}]{sterile}
\bibinfo{note}{A non-zero value for $\mathcal{A}_{NC}$ might be evidence for
  sterile neutrinos}.

\bibitem[{\citenamefont{{The SNO Collaboration}}(2000)}]{NIM}
\bibinfo{author}{\bibnamefont{{The SNO Collaboration}}},
  \bibinfo{journal}{Nucl. Instr. and Meth.} \textbf{\bibinfo{volume}{A449}},
  \bibinfo{pages}{172} (\bibinfo{year}{2000}).

\bibitem[{bib()}]{bib:companion}
\bibinfo{note}{Details of SNO response functions are available from
the SNO web site: http://sno.phy.queensu.ca}


\bibitem[{\citenamefont{{M. Dragowsky {\em et~al.}}}(2002)}]{N16}
\bibinfo{author}{\bibnamefont{{M. Dragowsky {\em et~al.}}}},
  \bibinfo{journal}{Nucl. Instr. and Meth.} \textbf{\bibinfo{volume}{A481}},
  \bibinfo{pages}{284} (\bibinfo{year}{2002}).

\bibitem[{\citenamefont{Fukuda {\em et~al.}}(2001)}]{bib:SuperK}
\bibinfo{author}{\bibfnamefont{S.}~\bibnamefont{Fukuda}} \bibnamefont{{\em et~al.}},
  \bibinfo{journal}{Phys. Rev. Lett.} \textbf{\bibinfo{volume}{86}},
  \bibinfo{pages}{5651} (\bibinfo{year}{2001}).

\bibitem[{bib()}]{Robustsig}
\bibinfo{author}{\bibnamefont{{J.~N. Bahcall, P. Krastev, and A.~Y.
  Smirnov}}}, \bibinfo{journal}{Phys. Rev. D}
  \textbf{\bibinfo{volume}{62}}, \bibinfo{pages}{093004}
  (\bibinfo{year}{2000});
\bibinfo{author}{\bibnamefont{{M. Maris and S.~T.
  Petcov}}}, \bibinfo{journal}{Phys. Rev. D}
  \textbf{\bibinfo{volume}{62}}, \bibinfo{pages}{093006}
  (\bibinfo{year}{2000});
\bibinfo{author}{\bibnamefont{{J.~N. Bahcall, M.~C. Gonzalez-Garcia, and C.
  Pe\~na-Garay}}}, (\bibinfo{year}{2002}), \eprint{hep-ph/0111150 v2.}

\bibitem[{bib()}]{bib:MSW}
\bibinfo{author}{\bibnamefont{{S.~J. Parke}}}, \bibinfo{journal}{Phys. Rev. Lett.}
  \textbf{\bibinfo{volume}{57}}, \bibinfo{pages}{1275}
  (\bibinfo{year}{1986});
\bibinfo{author}{\bibnamefont{{S.~T.
  Petcov}}}, \bibinfo{journal}{Phys. Rev. B}
  \textbf{\bibinfo{volume}{200}}, \bibinfo{pages}{373}
  (\bibinfo{year}{1988});
\bibinfo{author}{\bibnamefont{{G.~L. Fogli and E.
  Lisi}}}, \bibinfo{journal}{Astropart. Phys.}
  \textbf{\bibinfo{volume}{3}}, \bibinfo{pages}{185}
  (\bibinfo{year}{1995});
\bibinfo{author}{\bibnamefont{{E. Lisi, A. Marrone, D. Montanino, A.
  Palazzo, and S.~T. Petcov}}}, \bibinfo{journal}{Phys. Rev. D}
  \textbf{\bibinfo{volume}{63}}, \bibinfo{pages}{093002}
  (\bibinfo{year}{2001}).


\bibitem[{\citenamefont{Ortiz {\em et~al.}}(2000)}]{bib:8B}
\bibinfo{author}{\bibfnamefont{C.~E.} \bibnamefont{Ortiz}}
  \bibnamefont{{\em et~al.}}, \bibinfo{journal}{Phys. Rev. Lett.}
  \textbf{\bibinfo{volume}{85}}, \bibinfo{pages}{2909} (\bibinfo{year}{2000}).

\bibitem[{\citenamefont{Nakamura {\em et~al.}}(2002)}]{bib:xsec}
\bibinfo{author}{\bibfnamefont{S.}~\bibnamefont{Nakamura}} \bibnamefont{{\em et~al.}}
  (\bibinfo{year}{2002}), \eprint{nucl-th/0201062}.

\bibitem[{\citenamefont{{J.~N. Bahcall, M.~H. Pinsonneault and S.
  Basu}}(2001)}]{bib:SSM1}
  \bibinfo{author}{\bibnamefont{{J.~N. Bahcall, H.~M. Pinsonneault and
  S. Basu}}}, \bibinfo{journal}{Astrophys. J}
  \textbf{\bibinfo{volume}{555}}, \bibinfo{pages}{990}
  (\bibinfo{year}{2001}).

\bibitem[{\citenamefont{Cleveland {\em et~al.}}(1998)}]{bib:Cl}
\bibinfo{author}{\bibfnamefont{B.~T.} \bibnamefont{Cleveland}}
  \bibnamefont{{\em et~al.}}, \bibinfo{journal}{Astrophys. J}
  \textbf{\bibinfo{volume}{496}}, \bibinfo{pages}{505} (\bibinfo{year}{1998}).

\bibitem[{\citenamefont{Abdurashitov {\em et~al.}}(1999)}]{bib:Sage}
\bibinfo{author}{\bibfnamefont{J.~N.} \bibnamefont{Abdurashitov}}
  \bibnamefont{{\em et~al.}}, \bibinfo{journal}{Phys. Rev. C}
  \textbf{\bibinfo{volume}{60}}, \bibinfo{pages}{055801}
  (\bibinfo{year}{1999}).

\bibitem[{\citenamefont{Abdurashitov {\em et~al.}}(2002)}]{bib:Sage3}
\bibinfo{author}{\bibfnamefont{J.~N.} \bibnamefont{Abdurashitov}}
  \bibnamefont{{\em et~al.}}  (\bibinfo{year}{2002}), \eprint{astro-ph/0204245}


\bibitem[{\citenamefont{Altmann {\em et~al.}}(2000)}]{bib:Gallex}
\bibinfo{author}{\bibfnamefont{M.}~\bibnamefont{Altmann}} \bibnamefont{{\em et~al.}},
  \bibinfo{journal}{Phys. Lett. B} \textbf{\bibinfo{volume}{490}},
  \bibinfo{pages}{16} (\bibinfo{year}{2000}).

\bibitem[{\citenamefont{Hampel {\em et~al.}}(1999)}]{bib:GALLEX2}
\bibinfo{author}{\bibfnamefont{W.}~\bibnamefont{Hampel}} \bibnamefont{{\em et~al.}},
  \bibinfo{journal}{Phys. Lett. B} \textbf{\bibinfo{volume}{447}},
  \bibinfo{pages}{127} (\bibinfo{year}{1999}).

\bibitem[{\citenamefont{Cattadori {\em et~al.}}(2001)}]{bib:Gallex3}
\bibinfo{author}{\bibfnamefont{C.~M.}~\bibnamefont{Cattadori}}
\bibnamefont{{\em et~al.}}, in
  \emph{\bibinfo{booktitle}{Proceedings of the TAUP 2001 Workshop}}, 
  (\bibinfo{year}{September 2001}), Assergi, Italy.


\end{thebibliography}
\end{document}